\documentclass{article}

\usepackage[numbers]{natbib}

\usepackage{neurips_2022}

\usepackage{graphicx} 

\usepackage[utf8]{inputenc} 
\usepackage[T1]{fontenc}    
\usepackage{hyperref}       
\usepackage{url}            
\usepackage{booktabs}       
\usepackage{amsfonts}       
\usepackage{nicefrac}       
\usepackage{microtype}      
\usepackage{xcolor}         

\title{Rethinking Knee Osteoarthritis Severity Grading: A Few Shot Self-Supervised Contrastive Learning Approach}

\author{ Niamh Belton$^{1,2}$ \hspace{0.5cm}  Misgina Tsighe Hagos$^{1,3}$  \hspace{0.5cm} Aonghus Lawlor$^{3,4}$ \hspace{0.5cm} Kathleen M. Curran$^{1,2}$   \\
 \\ 
   $^{1}$Science Foundation Ireland Centre for Research Training in Machine Learning \\
    $^{2}$School of Medicine, $^{3}$School of Computer Science, University College Dublin \\ 
    $^{4}$Insight Centre for Data Analytics, University College Dublin, Dublin, Ireland \\
  \vspace{-0.9cm}
}

\begin{document}

\maketitle

\begin{abstract}
Knee Osteoarthritis (OA) is a debilitating disease affecting over 250 million people worldwide. Currently, radiologists grade the severity of OA on an ordinal scale from zero to four using the Kellgren-Lawrence (KL) system. Recent studies have raised concern in relation to the subjectivity of the KL grading system, highlighting the requirement for an automated system, while also indicating that five ordinal classes may not be the most appropriate approach for assessing OA severity. This work presents preliminary results of an automated system with a continuous grading scale. This system, namely SS-FewSOME, uses self-supervised pre-training to learn robust representations of the features of healthy knee X-rays. It then assesses the OA severity by the X-rays' distance to the normal representation space. SS-FewSOME initially trains on only \lq{}few\rq{} examples of healthy knee X-rays, thus reducing the barriers to clinical implementation by eliminating the need for large training sets and costly expert annotations that existing automated systems require. The work reports promising initial results, obtaining a positive Spearman Rank Correlation Coefficient of 0.43, having had access to only 30 ground truth labels at training time.

\end{abstract}

\section{Introduction}
Knee Osteoarthritis (OA) is a degenerative joint disease that affects over 250 million of the world's population \cite{kneestats}. Presently, the Kellgren-Lawrence (KL) system \cite{klscale} is the standard OA grading system consisting of five ordinal classes from grade zero to four where grade zero is healthy and grade four is severe OA. The subjectivity of the KL grading has been an ongoing subject of concern \cite{kl_concern}, highlighting the requirement for an automated system, while also indicating that five ordinal classes may not be the most appropriate approach for assessing OA severity. This work outlines SS-FewSOME, an automated continuous grading system for knee OA. It uses self-supervised pre-training to learn robust representations of the features of healthy knee X-rays and assesses the OA severity of an X-ray based on its distance in representation space to the centre of the normal representation space. This is the core concept that forms the basis of several anomaly detection techniques such as DeepSVDD \cite{deepsvdd}, Patchcore \cite{patchcore} and FewSOME \cite{fewsome}, where data samples that are sufficiently distant from the normal representations are deemed \lq{}anomalous\rq{}.

Although several automated grading systems based on Convolutional Neural Networks (CNN) \cite{3mod,chen,inter,lstm,multiscale} have shown promising results, these approaches require large datasets consisting of thousands of X-rays for training, along with ground truth OA severity labels from experts which is a tedious and costly process that is subject to annotator variability. SS-FewSOME initially trains on \lq{}few\rq{} examples of healthy knee x-rays and thus, it eliminates the aforementioned challenges and reduces the barriers to clinical implementation of Machine Learning based automated systems.

\begin{figure}  \includegraphics[width=\textwidth]{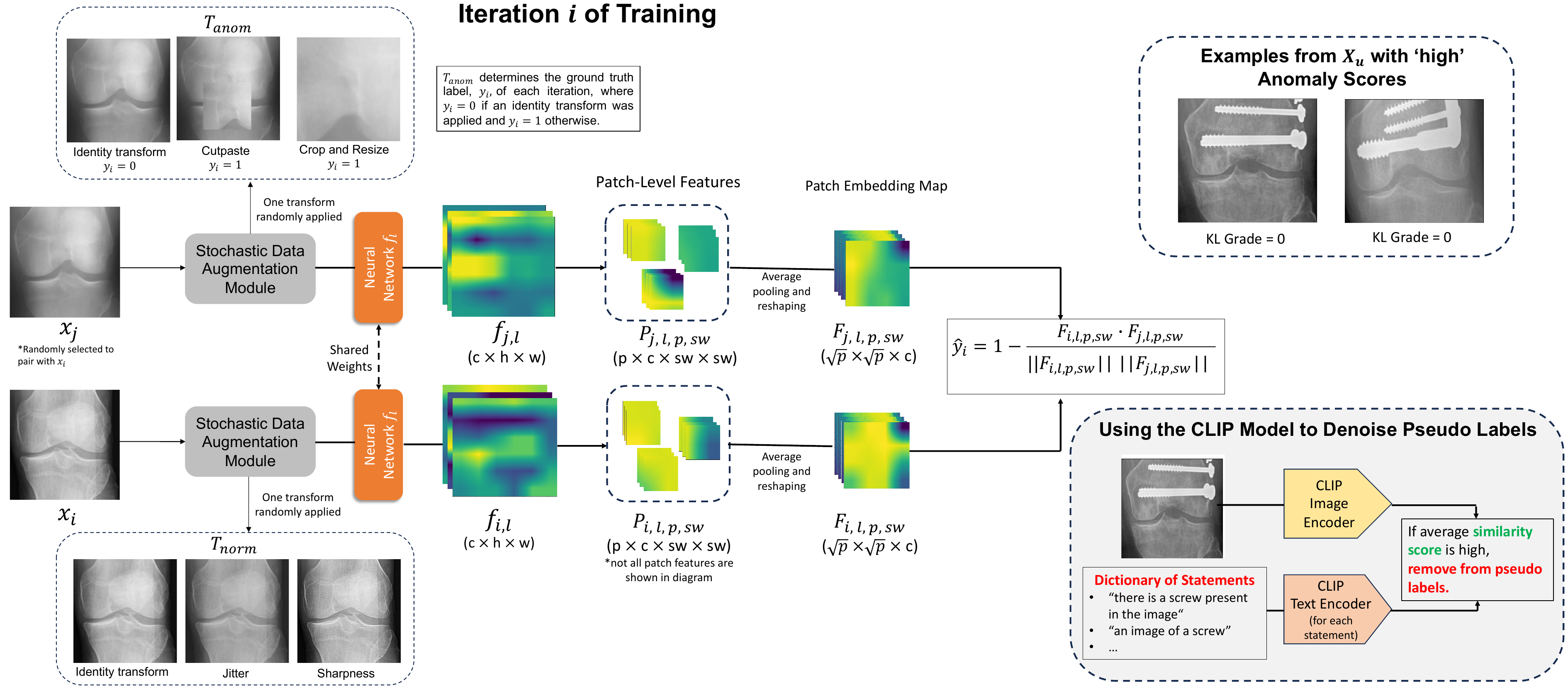}
  \caption{The figure visualises the self-supervised pre-training of SS-FewSOME as outlined in the text. The figure also shows two example X-rays with \lq{}high\rq{} anomaly scores due to the presence of metal in the X-ray. As shown in the bottom right of the figure, the CLIP model, along with a dictionary of statements can be used to identify such X-rays and remove them from the pseudo labelled X-rays before model retraining.}
\vspace{-0.3cm}

  \label{sense}
\end{figure}

\section{Methods}

\vspace{-0.2cm}

FewSOME was selected for the basis of this analysis
due to its performance in the few shot setting. Due to the success of self-supervised techniques in representation learning \cite{simclr,csi,spot}, FewSOME was extended to SS-FewSOME which includes a Stochastic Data Augmentation (SDA) module to enable self-supervised training. A dataset originally obtained from the Osteoarthritis Initiative (OAI) \cite{oai} containing 
bilateral Posterior Anterior fixed flexion X-ray images \cite{dataset} was used for this analysis.

\textbf{Self-Supervised Pre-Training} \hspace{0.2cm} A dataset $X_N$ of size $N$ consists of solely nominal images of healthy X-rays. In iteration $i$ of training, a data sample $x_{i} \in X_N$ is paired with a randomly selected data sample $x_j \in X_N$. The original FewSOME implementation relies on ImageNet \cite{imagenet} pretrained weights and an \lq{}anchor\rq{} for learning compact representations of the normal class where the ground truth label, $y$ is equal to zero for the duration of training. This work extends this method to use self-supervised training where both data samples, $x_i$ and $x_j$ are input into the Stochastic Data Augmentation (SDA) module. The SDA module consists of two sets of transforms, $T_{norm}$ and $T_{anom}$. The set of transforms $T_{norm}$ are applied only to $x_i$ and they consist of weak, global augmentations that aim to generate additional X-rays whose representation are within the hyper-sphere of the  normal representation space. For this analysis, $T_{norm}$ consists of the identity function (i.e. no transformation), applying jitter to the image and adjusting the sharpness and brightness. The set of transforms $T_{anom}$ are applied only to $x_j$ and they consist of the identity function and strong augmentations, random cropping followed by resizing and Cutpaste \cite{cutpaste}. These transforms were shown through experimentation to transform the image's representation to outside of the normal representation space. The SDA determines the ground truth label with $y_i=0$ if the identity transform was applied to $x_j$ and $y_i=1$ otherwise.


Following the SDA, the first $l$ layers of an AlexNet initialised with ImageNET pre-trained weights, denoted as $f_{l}$, are used to transform the input space, $X_{N}$ to the representation space, $f_l(X_{N}) \in \mathbb{R}^{c \times h \times w}$, where $c$ is the number of channels, $h$ is the height and $w$ is the width of the extracted feature maps 
and $l$ is set to $5$. The later layers are removed as they are biased towards natural image classification. To improve the sensitivity of the model to localised anomalies such as osteophytes, $f_l(x_i)$, denoted as $f_{i, l}$, is converted to patch level features. A sliding window of size $sw \times sw$ with $stride =1$ converts $f_{i, l} \in \mathbb{R}^{c \times h \times w}$ to patch level features, $P_{i, l, p, sw} \in \mathbb{R}^{p \times c \times sw \times sw}$ where $p$ is the number of patches. Feature aggregation is performed for each patch $p$ by average pooling and the resultant tensor is reshaped to create a patch embedding map, $F_{i, l, p, sw} \in \mathbb{R}^{\sqrt{p} \times \sqrt{p} \times c}$. 

This method exploits that X-rays are spatially very similar and therefore, the cosine similarity is calculated only between patches in $F_{i, l, p, sw}$ and $F_{j, l, p, sw}$ that have the same coordinates. The loss function, Binary Cross Entropy was calculated based on one minus the patch cosine similarity scores and the ground truth value, $y_i$ (where the value of $y_i$ was determined by the SDA module). %


\textbf{Pseudo Labelling and Denoising with CLIP} \hspace{0.2cm}
This method is extended for the setting where additional X-rays are available at training time but they are without expert ground truth labels. Given an unlabelled dataset of X-rays, $X_u$ of size $u$ where $X_u \cap X_N = \emptyset$, the patch embedding map, $F_{u, l, p, sw}$, for each data sample $x \in X_u$ is obtained, followed by the anomaly score which is equal to one minus the mean cosine similarity between $F_{u, l, p, sw}$ and  $F_{x, l, p, sw} \forall x \in X_N$. X-rays with high anomaly scores are assumed to be severe OA cases. The model can be retrained on these cases to improve the sensitivity to OA specific anomalies with the newly assigned pseudo label of $y=1$. 
However, knee X-rays often contain anomalies unrelated to the OA grade, a common one being the presence of screws and metal in the X-ray. The CLIP \cite{clip} model along with a dictionary of statements is used to denoise the pseudo labels. The dictionary consists of statements such as \lq{}there is a screw present in the image\rq{}. X-rays with significantly high similarity scores to the statements were removed. A denoised dataset $X_d \subset X_u$ of size $d$ then consists of X-rays that had a low similarity score to the statements and also a sufficiently high anomaly score. \lq{}High\rq{} anomaly scores have values greater than twice the average score between each pair of normal representations in the training data. 


\textbf{Retraining with Pseudo Labels} \hspace{0.2cm}
Retraining the model with pseudo labels is similar to the self-supervised pre-training with some key differences. The SDA is switched off and the value of $y$ is determined by where $x_j$ is sampled from. For example, in iteration $i$ of training, $x_i \in X_N$ is paired with a randomly selected data sample $x_j$ where $y_i=0$ if $x_j \in X_N$ and $y_i=1$ if $x_j \in X_d$. Additionally, as there is now more data available and therefore, the model is less prone to overfitting, the original AlexNet architecture is replaced with the larger VGG-16. The objective of the training remains the same, to obtain the patch embedding map for both $x_i$ and $x_j$, calculate one minus the patch cosine similarity scores and compare this to the ground truth $y_i$.

\vspace{-0.25cm}
\section{Results and Conclusion}
\vspace{-0.25cm}

Table \ref{tab1} shows the results on a test set consisting of 1,526 X-rays before the model was retrained on pseudo labels. The table shows the Spearman Rank Correlation Coefficient (SRCC) for competing methods DeepSVDD and Patchcore, the original implementation of FewSOME and the proposed SS-FewSOME with $N=30$ for all methods, meaning the model trains on just 30 examples of healthy knee X-rays. SRCC measures the correlation between the model output scores and the ground truth KL grades. The results are averaged over five seeds with $X_N$ being randomly sampled from the dataset each time to demonstrate the model's robustness to the contents of the training data. SS-FewSOME outperforms competing methods, demonstrating its potential for use as an automated continuous OA grading system. The table also highlights the model's performance on the task of detecting severe OA cases (KL Grade 4) in terms of Area Under the Curve (AUC).

The model was then used to assign pseudo labels on the remaining training data and the X-rays with \lq{}high\rq{} anomaly scores were considered for retraining. Prior to denoising the pseudo labels with CLIP, the percentage of grade zeros with high anomaly scores had an average value of 10.7\%. This decreased to 0.7\% after denoising the labels. The model was then retrained with $d=55$ (average) denoised pseudo labels. The results in table \ref{tab1} show that the model is sensitive to noisy pseudo labels as the model performance is similar when trained on no pseudo labels and when trained with pseudo labels that were not denoised, both models scoring a SRCC of 0.35. However, by retraining with the denoised pseudo labels, the model's performance has increased to an SRCC value of $0.43$, and it can now detect severe cases of OA with an AUC of 91.2\%. This is a promising result given a recent study found the interrater reliability between experts using the KL system can be as low as 0.51 \cite{agreement}.

This preliminary work has demonstrated SS-FewSOME's ability as an automated continuous OA grading system whilst having a low data and annotation requirement, thus reducing the barriers to clinical implementation. The next steps will focus on further developing the pseudo labelling technique and performing a clinical evaluation of the proposed continuous grading system.

\vspace{-0.5cm}

\begin{table*}
\small
  \centering
  \begin{tabular}{c|cccc|cc}  
    Metric & DSVDD & Patchcore & FSOME & SS-FSOME  & \begin{tabular}{@{}c@{}}SS-FSOME \\ $_{psuedo} $\end{tabular}  & \begin{tabular}{@{}c@{}}SS-FSOME \\ $_{pseudo \times CLIP}$\end{tabular} \\
    \hline
 
    $AUC_{g=4}$ & $65.0\pm{}1.5$  & $84.4\pm{}1.4$  & $67.1\pm{}2.1$ & $\bold{86.6\pm{}2.2}$ & $85.5\pm{}6.9$ & $\bold{91.2\pm{}2.5}$
    \\

    SRCC & $0.10\pm{}0.03$  & $0.22\pm{}0.01$ & $0.13\pm{}0.02$ & $\bold{0.35\pm{}0.02}$  & $0.35\pm{}0.12$ & $\bold{0.43\pm{}0.02}$
    \\

  \end{tabular}
  \caption{Average AUC for detecting severe OA ($grade=4$) in \% and average Spearman Rank Correlation Coefficient (SRCC) calculated over five seeds for methods DeepSVDD (DSVDD), Patchcore, FewSOME (FSOME), SS-FewSOME (SS-FSOME), SS-FewSOME trained on psuedo labels (SS-FSOME$_{pseudo}$) and SS-FewSOME trained on denoised pseudo labels (SS-FSOME$_{pseudo \times CLIP}$).}

  \vspace{-0.2cm}
  \label{tab1}
\end{table*}

\section*{Acknowledgements}
This work was funded by Science Foundation Ireland through the SFI Centre for Research Training in Machine Learning (Grant No. 18/CRT/6183). This work is supported by the Insight Centre for Data Analytics under Grant Num- ber SFI/12/RC/2289 P2.

\bibliographystyle{plainnat}

\bibliography{ref}

\end{document}